\documentstyle[prl,aps,epsf,multicol]{revtex}
\begin{document}

\title{
Superconductivity mediated by charge fluctuations in   
layered molecular crystals
}
\draft
\author{Jaime Merino\cite{email0} and Ross H. McKenzie }
\address{Department of Physics, University of Queensland,
Brisbane 4072, Australia}
\date{\today}
\maketitle
\begin{abstract}
 There is no consensus about
   the mechanism of the superconductivity or
 the pairing symmetry 
for layered molecular crystals.
Applying slave-boson theory to an extended Hubbard model 
we show that 
for  the $\theta$ and $\beta$'' crystal structures
the superconductivity is mediated
by charge fluctuations and the order parameter has
d$_{xy}$ symmetry. This is in contrast to
the $\kappa$-(BEDT-TTF)$_2$X family,
for which theoretical calculations give
superconductivity mediated by spin fluctuations
and with d$_{x^2-y^2}$ symmetry.
This is the simplest model that can describe
the competition between metallic, superconducting,
insulating, and charge ordered phases that occurs
in the $\theta$ and $\beta$'' materials.
We predict several materials that 
should become superconducting under pressure.
\\
\end{abstract}
%\pacs{PACS numbers: 71.27.+a, 71.10.Fd, 74.70.Kn}

\begin{multicols}{2}
%\columnseprule 0pt
\narrowtext

The issue of the interplay of superconductivity,
magnetism, and charge ordering is relevant
to a wide range of strongly correlated electron
materials. Examples include the copper-oxide
(high-temperature) superconductors \cite{Tranquada}, colossal
magnetoresistance materials \cite{MoriS},
 heavy fermion
compounds \cite{Mathur},
vanadium oxides \cite{ueda},
 and organic molecular crystals\cite{ishiguro,science,mckenzie}.
In particular for the cuprate superconductors
there is controversy about the relative
importance of charge fluctuations
(associated with dynamical 
"stripes") and antiferromagnetic spin
fluctuations (associated with the 
 Mott insulator which occurs when there
is an average of one electron or hole for every lattice site).
For some heavy fermion compounds 
recent experiments support the idea that the
superconductivity is mediated by spin fluctuations \cite{Mathur}.

The family $\kappa$-(BEDT-TTF)$_2$ X \cite{greek} of 
molecular crystals have
similarities to the cuprates\cite{science} including
the proximity of superconductivity to a Mott insulator
in the phase diagram. Although there is an average of
half a hole per molecule the necessary condition
of one hole per lattice site is met because
the molecules are paired up (dimerized) within
the $\kappa$- type crystal structure.
Theoretical calculations\cite{Schmalian}
suggest          that the superconductivity has
d$_{x^2-y^2}$ symmetry (as in the cuprates) and is
mediated by antiferromagnetic spin fluctuations.
However, there is controversy about whether experiments
support this \cite{elsinger}.
In this Letter, we show theoretically that 
the organic superconductors listed in Table \ref{table1}
are quite different from the $\kappa$- type materials and
the superconductivity is mediated
by charge fluctuations and has
d$_{xy}$ symmetry. 
Our results may also be relevant
to the recent discovery that the quasi-one-dimensional
vanadium bronzes
 $\beta'$-Cu$_{0.65}$V$_2$O$_5$
 and
 $\beta$-Na$_{0.33}$V$_2$O$_5$
(which are  close to a charge ordered insulator) becomes
superconducting under pressure \cite{ueda}.
Previously, it was suggested by Scalapino, Loh, and
Hirsch \cite{scalapino} that spin fluctuations could mediate
d$_{x^2-y^2}$ pairing and charge fluctuations could
mediate d$_{xy}$ pairing.

The materials                 considered here
consist of layers of donor molecules
[e.g., BEDT-TTF=bis-(ethylenedithiatetrathiafulvalene)] alternating 
with insulating layers of anions [e.g., X=SF$_5$CH$_2$CF$_2$SO$_3$].
For the $\theta$ and $\beta$'' 
crystal structures the donor molecules are not
dimerized and so non-interacting electron models
(band structure calculations)  predict
a metallic state due to a band which 
is one quarter filled with holes.
However, some of these materials are insulators at low
temperatures. Mott insulators
(resulting from the Coulomb repulsion between
electrons on a single site) only occur for a half-filled band.
However,
the localization of charge (and associated
insulating behavior) could result from charge ordering
due to the Coulomb repulsion between electrons
on neighbouring sites.
Indeed, such charge ordering is observed in
some of these materials and is reflected
in a disproportion of charge between neighbouring
donor molecules
(see Ref.\cite{mckenzie} for a brief review of how this is
determined experimentally).
Depending on the anion, temperature and pressure,
the materials can be either a charge ordered insulator, a metal, or 
a superconductor. (A schematic phase diagram
of the $\theta$ materials is shown in Fig. 1 of Ref.\cite{mckenzie}).
The common feature of the superconductors listed
in Table I is that in the phase diagram
the superconductivity occurs in
close proximity to the insulating and/or charge 
ordered phase. Five of the superconductors
have the very unusual property that as the
temperature decreases the resistivity is increasing
before entering the superconducting phase.

\begin{figure}
%\centerline{\epsfxsize=15cm \epsfbox{fig1.eps}}
\centerline{\epsfxsize=9cm \epsfbox{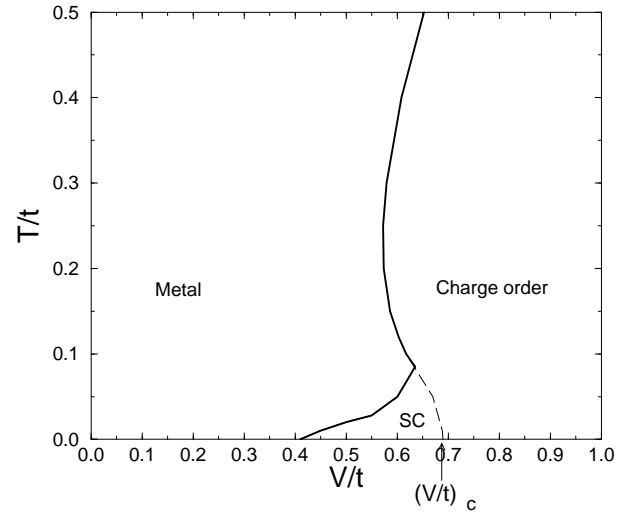}}
\caption{Phase diagram showing 
competition between metallic, superconducting (SC), and charge 
ordered phases. 
 The symmetry of the Cooper pairs in the
superconducting phase is d$_{xy}$.
The phase diagram is for 
 the extended Hubbard model
 (defined by the Hamiltonian (\ref{ham}))
with an average of one hole per two molecules
(a quarter-filled band) in the limit of
infinite Coulomb repulsion energy $U$ for two holes on
the same molecule.
This is the simplest model Hamiltonian which can describe
the organic superconductors listed in Table I.
The vertical axis is the ratio of the temperature $T$ to
the intermolecular hopping integral $t$.
The horizontal axis is the ratio of the nearest-neighbour
Coulomb repulsion energy $V$ between electrons
on neighbouring molecules to $t$.
%This phase diagram was obtained from large-$N$ theory applied to 
% to next-to-leading order in 1/$N$, for $N=2$.
Note that the unconventional superconductivity
is found near the quantum critical point (at $(V/t)_c$)
separating the metallic and  charge ordered phases.
}                   
\label{fig1}
\end{figure}

The simplest possible strongly correlated electron model 
which can describe the competition among the above phases 
is an extended Hubbard model at quarter-filling on a square 
lattice\cite{mckenzie}. The Hamiltonian is 
\begin{eqnarray}
H &=& t \sum_{<ij>,\sigma}
(c^\dagger_{i \sigma} c_{j \sigma} +
c^\dagger_{j \sigma} c_{i \sigma})
+ U \sum_{i} n_{i\uparrow} n_{i\downarrow}
\nonumber \\
&+& V \sum_{<ij>} n_i n_j - \mu \sum_{i \sigma} n_{i \sigma} %\\
%\nonumber \\
\label{ham}
\end{eqnarray}          
where the operator
$c^\dagger_{i \sigma}$ creates an 
electron in site $i$ with spin $\sigma$. $t$ is the amplitude for electrons 
to go from one site to a nearest-neighbours one, $V$ is a nearest-neighbours 
Coulomb interaction, and $U$ is the electron-electron interaction at 
a given site. $\mu$ is the chemical potential.
%A simpler version of hamiltonian (\ref{ham}) 
%for electrons without spin, the so-called Cullen-Callen model\cite{Cullen},
%captures the essential physics involved in the
%metal-insulator transition found in Magnetite.
A further simplification on model (\ref{ham}) can 
be made considering the fact that  
$U >> V, t$, so we can fix $U = \infty$. 
We have previously studied charge ordering within 
this model\cite{mckenzie} and here we briefly summarize the main results.
For $V >> t$, the model has                    
an insulating phase with checkerboard charge ordering ({\it i.e.},
the wavevector associated with charge modulation is $(\pi,\pi)$,
see Fig. 3 in Ref.\cite{mckenzie}).
The spins in the charge ordered state are antiferromagnetically
coupled due to a fourth order ring-exchange process.           
For $V/t < (V/t)_c \approx 0.69$ we find a metallic phase 
with homogeneous charge density while for $V/t > (V/t)_c $
the system becomes charge ordered.
Hence, we find a quantum phase transition from a metallic 
phase to a charge ordered phase with a quantum critical point
at $(V/t)_c$. It should be stressed that this charge ordering 
instability is not associated with nesting of the Fermi 
surface since at quarter-filling the Fermi surface diameter is much smaller
than the length of the $(\pi, \pi)$ wave vector \cite{note}.
We now give a brief outline of
our new theoretical calculations which show how charge fluctuations near
this quantum critical point produce superconductivity.
Our main results are summarised in Fig. \ref{fig1}.

First, we extend the SU(2) spin symmetry of (\ref{ham}) to 
SU($N$) by allowing the index $\sigma$ to run from 1 to $N$.
We have used slave-boson theory\cite{Barnes} combined with  a 
1/$N$ expansion\cite{Read}, where $N$ 
is considered to be large.
This type of approach has previously been applied to a wide range of
strongly correlated electron systems including
the Kondo model\cite{Barnes,Read,Hewson},
 heavy fermions\cite{Burdin},
 the Hubbard model\cite{Ruckenstein},
and models for the cuprate superconductors\cite{Kotliar,Castellani}.
It has been successful in describing the physics
of the Kondo effect\cite{Hewson} and, for the Hubbard model, 
it can describe the Mott-Hubbard metal-insulator transition\cite{Ruckenstein}.
The mean-field theory in the slave bosons corresponds to the
Gutzwiller approximation\cite{Ruckenstein}, and so the 1/$N$ expansion
provides a systematic method to calculate corrections to this approximation.    We briefly outline the main steps of the approach;
details can be found elsewhere\cite{Read,Kotliar,mckenzie}.
The condition $U \rightarrow \infty$  precludes 
doubly-occupied sites and so it is convenient to introduce the following 
representation\cite{Barnes} for the electron operators:
$c^{\dagger}_{i\sigma}=f^{\dagger}_{i \sigma} b_i$,
where $f^{\dagger}_{i \sigma}$ represents a fermion at site $i$ which 
carries spin $\sigma$, and $b_i$, is a boson associated with the 
electron charge located at site $i$. 
We impose the constraint that either a fermion or a boson can
be at each lattice site,   
$f^{\dagger}_{i \sigma} f^{\dagger}_{i \sigma} + b^+_i b_i =N/2$,
by introducing a Lagrange multiplier, $\lambda_i$ at each lattice site.
We expand to the next-to-leading order corrections in 1/$N$,
arriving at a Hamiltonian which describes an effective
coupled fermion-boson problem: $H  =  H^{f}+H^{b}+H^{f-b}$.
%\begin{eqnarray}
%$H & = & H^{f}+H^{b}+H^{int} 
%\nonumber \\
%H^{f} & = & \sum_{{\bf k} \sigma } \epsilon_{\bf k} f^{\dagger}_{{\bf k} \sigma} 
%f_{{\bf k} \sigma}\nonumber  \\
%H^{b} &= & N \sum_{{\bf q} \alpha \beta} \delta R_{\alpha}({\bf q}) 
%A_{\alpha \beta} ({\bf q}) \delta R_{\beta}(-{\bf q}) \nonumber  \\%
%H^{int} & = & \sum_{{\bf k q} \sigma} f^{\dagger}_{{\bf k + q} \sigma}
%\Lambda_{\alpha} ({\bf k},{\bf q}) 
%f_{{\bf k} \sigma} \delta R_{\alpha}({\bf q})
%\nonumber \\
%\label{hamB}
%\end{eqnarray}
$H^{f}$ is the result of taking the average of the boson fields:
$b= \langle b_i \rangle$ and $\lambda= \langle \lambda_i 
\rangle$, and describes fermions moving in a renormalized band 
with energy dispersion
given by $\epsilon_{\bf k} = {- t b^2 \over N} T_{\bf k} + \lambda - \mu + 4 V {\langle n \rangle \over N} $, with $T_{\bf k}=2( \cos(k_x) + \cos(k_y) )$,
being the Fourier transform of the hopping 
operator in units of the nearest-neighbour hopping $t$.                 
$\langle n \rangle $ is the average occupation number of the electrons
at each lattice site.
The effect of Coulomb interactions is then two-fold: (i)
renormalization of the free electron energies by a factor
 $b^2=N/2- \langle n \rangle$,
(ii) upward shift in the position of the band which is given by
$\lambda =\sum_{\bf k} f(\epsilon_{\bf k})(t T_{\bf k} + 4 V)$. 
 For $N=2$ and a quarter filled-band, $\langle n \rangle ={1 \over 2}$,
 and so  the effective mass of the quasiparticles is 
enhanced by a factor of $m^*/m=1/b^2=2$.

The bosonic part of the Hamiltonian, $H^b$, describes the dynamical and
spatial fluctuations about the mean-field solution.
The field $\lambda_i$ describes fluctuations in the no-double 
occupancy constraint at each lattice site.
The real charge fluctuations are described by $b_i$.
%The bosonic fields $\delta \hat{R}$ are given by $ 
%\delta \hat{R}({\bf q}) = (\delta r({\bf q}), \delta \lambda ({\bf q}) ) 
%$ and the fermion-boson vertices are defined as
%\begin{equation}
%\hat{\Lambda}({\bf k}, {\bf q}) = ( {-t
%b^2 \over N} ( T_{\bf k} + T_{\bf k +q} ) - {2 b^2 V \over N}
%V_{\bf q}, i) 
%\end{equation}
%where $V_{\bf k}=2( \cos(k_x) + \cos(k_y) $ 
%is the Fourier transform of the Coulomb interaction
%operator, and the $A$ matrices representing the free bosonic part
%of the hamiltonian is 
%\begin{equation}
%\hat{A}({\bf q})= \left( \begin{array}{cc}
%{2 b^2 \lambda \over N} -{2 b^2 t \over N} \sum_{\bf k} (T_{\bf k-q}
%+ {V \over t} V_{\bf k} ) f(\epsilon_{\bf k} ) & i {2 b^2 \over N} \\
%i {2 b^2 \over N} & 0 \\ \end{array}\right)   
%\end{equation}
These boson fields propagate according to, $\hat{D^0}({\bf q}, i \nu_n)$, 
where ${\bf q}$ is the momentum of the boson and $\nu_n= 2 \pi n T$ 
is a Matsubara frequency with $T$ being the temperature and $n$ an
integer number. Finally, $H^{f-b}$, couples the fermions and the bosons
so that they propagate according to: $ \hat{D}({\bf q}, \nu_n)={1 \over N} 
(\hat{D^0}({\bf q}, \nu_n)^{-1} - \hat{\pi}({\bf q}, \nu_n) )^{-1} $
where $\hat{\pi}({\bf q}, \nu_n )$ is the self-energy of the bosons.
%The solution to this effective fermion-boson problem   
%\begin{equation}
%\hat{D}({\bf q}, \nu_n)={1 \over N} 
%(\hat{A} + \hat{\Pi}({\bf q}, \nu_n) )^{-1}
%\end{equation}
%with the polarization propagators given by
%\begin{equation}
%\Pi_{\alpha \beta}({\bf q}, \nu_n)=N \sum_{\bf k}
%{ f(\epsilon_{\bf k +q}) - f(\epsilon_{\bf k} ) \over 
%\epsilon_{\bf k + q} - \epsilon_{\bf k} - i \nu_n } 
%\Lambda_{\alpha}({\bf k},{\bf q}) \Lambda_{\beta} ({\bf k}, {\bf q})
%\label{Pi}
%\end{equation}
These bosons originate from the electron-electron
repulsion and produce  interaction
between the quasi-particles. In particular,
they can induce Cooper pairing of quasi-particles.
 This is in analogy to the pairing of electrons due
 to phonons in elemental metals.  
However, the mechanism of      the pairing in the present case 
is the    charge fluctuations.
%The quasiparticle scattering amplitude between the quasiparticles
%is given by
%\begin{eqnarray}
%&\Gamma&({\bf k}, {\bf -k}| {\bf k'}, {\bf -k'}) =
%\nonumber \\
%&-&\sum \Lambda_\alpha({\bf k}, {\bf k'})
%D_{\alpha \beta} ({\bf q= k - k'}, \nu=0) \Lambda_\beta({\bf -k}, {\bf -k'})
%\nonumber \\
%&+&\sum \Lambda_\alpha({\bf k}, {-\bf k'})
%D_{\alpha \beta} ({\bf q= k + k'}, \nu=0) \Lambda_\beta({\bf -k}, {\bf k'})
%\nonumber \\
%\label{scat}
%\end{eqnarray}
%where the first term in the right hand side represents the direct
%interaction, and the second one, the exchange processes between the
%quasiparticles in the Cooper pairs, and $\Lambda_{\bf \alpha}$ are
%the fermion-boson vertices\cite{mckenzie} associated with each of the
%boson fields denoted by $\alpha$.
%\section{Results}
The scattering amplitude in the particle-particle channel between 
a quasiparticle with momentum ${\bf k}$ and another with momentum 
${\bf -k}$ which scatter to ${\bf k'}$ and $-{\bf k'}$, 
is denoted $\Gamma({\bf q=k-k'})$. It can also be
understood as the effective potential between the quasiparticles
forming the Cooper pairs.  
%In Fig. \ref{fig2} we plot $\Gamma({\bf q})$,
%along the $q=q_x=q_y$ direction. 
A divergence in the effective interaction occurs at 
the charge ordering instability\cite{mckenzie,Castellani}.
We find that as the ratio $V/t$ is increased, the potential
varies its shape, developing singularities at $(\pm \pi, \pm \pi)$
when $V/t \rightarrow (V/t)_c \approx 0.69$, at zero temperature.
This signals the onset of checkerboard charge ordering.     

In order to look for superconducting instabilities near the 
charge-ordering instability we compute
Fermi surface averages\cite{Kotliar,scalapino} of 
the effective potential, $\Gamma({\bf q})$, 
weighted with the different cubic harmonics, into which the effective 
potential can be decomposed. 
These have different symmetries and they read         
%\begin{equation}
%\lambda_i= {1 \over (2 \pi)^2 }{ \int (d{\bf k}/|v_{\bf k}|) \int 
%(d {\bf k'}/|v_{\bf k'}|) g_i( {\bf k}) 
%\Gamma( {\bf k}, {\bf -k} | {\bf k'}, {\bf -k'}) g_i({\bf k'}) 
%\over \int (d k/|v_{\bf k}|) g_i({\bf k})^2   }
%\label{lambda}
%\end{equation}
\begin{eqnarray}
g_{s^*}({\bf k})=\cos(k_x)+\cos(k_y) 
\nonumber \\
g_{d_{x^2-y^2}}({\bf k})=\cos(k_x)-\cos(k_y)  
\nonumber \\
g_{d_{xy}}({\bf k})= \sin(k_x) \sin(k_y)  
\nonumber \\
g_{p_x}({\bf k})=\sin(k_x).
\label{armon}
\end{eqnarray}
From these Fermi surface averages, we find that there
is attraction between the quasiparticles forming Cooper
pairs with $d_{xy}$ symmetry at $V/t \geq 0.4$ and zero temperature. 
Cooper pairing with other symmetries are found to be repulsive for 
all values of $V/t$.
%This implies unconventional pairing with d$_{xy}$ symmetry near 
%the quantum critical point which separates the metallic from the 
%charge ordered state.
 This is in contrast to the 
d$_{x^2-y^2}$ symmetry found for 
$\kappa$-(BEDT-TTF)$_2$X within  renormalized
spin-fluctuation calculations\cite{Schmalian}. 
%Note that although up to $V/t= 0.6$, the effective potential is
%always repulsive, the couplings with d$_{xy}$ symmetry 
%are negative.                       
%The possibility of having superconducting pairing for
%a fully repulsive potential has been discussed in the
%context of the spin-fluctuation scenario\cite{Scalapino1}
%for the high-T$_c$ superconductors.

The finite-temperature phase diagram obtained
% within the large-$N$ approach in
%the slave-boson representation to order 1/$N$ with $N=2$ 
is shown   in Fig. \ref{fig1}. 
The line separating the metal from
the charge-ordered phase     is defined from the divergence at 
${\bf q} =(\pi, \pi)$ of the quasiparticle scattering amplitude,
$\Gamma(\bf q)$. The dashed line 
in Fig. \ref{fig1} is the extension of this  line, but we do not compute
the transition from the charge-ordered phase to the superconducting
phase. 
%Note            that the metal-insulator 
%transition shows re-entrant behaviour within the slave-boson 
%calculation used here \cite{Bulla}. 
As  the temperature is lowered it is possible to go from the charge ordered 
state directly into the superconducting phase.    
This re-entrant behaviour \cite{Bulla}
 might explain  the most unusual property 
 ($d \rho/d T < 0$)  
of five of the materials listed in Table \ref{table1}.
%Furthermore, from Fig. \ref{fig1}
%we can also explain the 'anomalous' behaviour of the resistivity observed
%in $\beta$''-(BEDT-TTF)$_2$Pd(CN)$_4$H$_2$O and $\beta$''-(BEDT-TTF)$_2$
%Pt(CN)$_4$H$_2$O, which are both insulating materials\cite{MoriT,MoriH}
%%at ambient pressure, with a minimum in the resistivity at
%T$_{MI}$=70 K and 120 K, respectively.
% At the critical pressure at which superconductivity
%occurs, both materials show an increasing resistivity with lowering temperature
%before becoming superconducting (note the negative derivative of the 
%resistivity above T$_c$ in some of the materials appearing in 
%Table \ref{table1}). 
%From our description, these materials are in 
%the charge ordered phase shown in Fig. \ref{fig1}, but close to 
%the superconducting phase. From Fig. \ref{fig1}, we observe that
%It is also consistent
%with the temperature dependence of the resistivity found in
%$\beta$''-(ET)$_2$ SF$_5$CH$_2$CF$_2$SO$_3$ which shows insulating
%behaviour\cite{Wang} with charge ordering for $T > 100 K$, and is
%metallic below this temperature before becoming superconducting at 5.2K.

The qualitative features of the phase diagram,
including the d$_{xy}$ pairing symmetry,
turn out to be {\it insensitive to the details
of the band structure and the type of charge ordering.}
First, we have changed the shape of the Fermi surface 
by introducing a next-nearest neighbour hopping     along one of 
the diagonals of the square lattice, $t'$ \cite{theta}. 
We find that varying the ratio, $t'/t$, in the range $0 \le t'/t \le 1$
changes the shape of the Fermi surface significantly (see Fig. 5 in 
Ref. \cite{mckenzie}) but does not 
destroy the d$_{xy}$ pairing instability. 
%Indeed, it only changes the critical
%lines shown in Fig. \ref{fig1} but qualitatively the 
%phase diagram is the same as in Fig. \ref{fig1}.                  
Second, introducing a next-nearest neighbours repulsion, $V'$, 
changes the momentum dependence of the scattering amplitude.       
Furthermore, for sufficiently large values of the ratio $V'/V$ the
singularities in $\Gamma({\bf q})$ shift from $(\pi, \pi)$ to 
$(0, \pm \pi)$ and $(\pm \pi, 0)$. 
This is because for $V'/V > 1$, it is energetically
more favourable to produce charge ordering either along the 
x or y-direction, rather than along both of them.
We find that there is still a quantum critical point, and superconducting 
pairing with d$_{xy}$ symmetry persists, although the strength of 
the effective interaction within the Cooper pairs decreases somewhat as 
compared to the square lattice case. 
%Recent optical conductivity experiments\cite{Wang1} in 
%$\theta$-(BEDT-TTF)$_2$RbZn(SCN)$_4$ show the opening of 
%an optical gap of about 300 cm$^{-1}$ below the metal-insulator transition
%temperature, $T < T_{MI} =190 K$.
%This gap appears along the stacking direction of the 
%molecules so that it is attributed to the presence of 
%horizontal arrays of charge below the metal-insulator transition
%temperature, $T < T_{MI} =190 K$. Within our theory, 
%this can be explained from the fact that $V'> V$. 
The robustness of the d$_{xy}$ superconducting instability
 can be understood from the fact that the charge-ordering 
instability is {\it not} associated with Fermi surface nesting. 
Furthermore, it means
our        results should also be applicable 
to the $\beta$'' materials for which the Fermi surface is
more complicated than for the $\theta$ materials.

{\it Proposed experimental tests.}
(i) %New superconductors.
We identify several materials that
% which at ambient pressure are not superconducting,
 might become superconducting under 
pressure. Pressure in the $\theta$ materials  decreases 
the hopping parameter $t$, driving the materials into the insulating phase
\cite{Mori}. 
%This has been observed\cite{Hanazato}, for example,
% in $\theta$-(BO)$_2$Cl(H$_2$O)$_3$.
Therefore, $\theta$ materials such as $\theta$-(BETS)$_2$Ag(CN)$_2$, 
$\theta$-(BETS)$_4$Cu$_2$(Cl)$_6$ and $\theta$-(BO)$_2$Cl(H$_2$O)$_3$, 
which are all metallic at ambient pressure, should become superconducting 
under pressure. These materials would then be located in 
the metallic side of our phase diagram (see Fig. \ref{fig1}), and pressure 
would increase $V/t$ driving them into a superconducting state 
with d$_{xy}$ symmetry before becoming insulating. 
(ii)% Raman scattering.
Polarisation-dependent Raman scattering should
be done in the superconducting state because
it can distinguish  $d_{xy}$ and $d_{x^2 - y^2}$ states \cite{raman}.
(iii)% NMR.
 Measurements of the nuclear magnetic resonance
relaxation rate and Knight shift should be done in the
metallic phase for the superconductors in Table I.
There should be no enhancement of the Korringa ratio.
This is in contrast to the 
large enhancements seen in
$\kappa$-(BEDT-TTF)$_2$X superconductors which are
close to the Mott insulator \cite{slichter}.

In conclusion, we have identified 11 molecular superconductors
which we predict to have pairing of $d_{xy}$ symmetry
due to charge fluctuations. This is based on a systematic
many-body calculation using slave boson theory
for an extended Hubbard model.
Materials with the $\theta$ and $\beta''$ crystal
structures are at quarter filling and are described
by quite different physics from $\beta$, $\kappa$, and
$\lambda$ structures which are essentially at
half filling.

 We thank E. Abrahams, G. Blumberg, O. C\'epas, P. Coleman,
H. Fukuyama,
A. Houghton, G. Kotliar, J.B. Marston, S. Mazumdar,
D.J. Scalapino, C. Slichter, and J. Wosnitza
 for very helpful discussions. This work was supported by 
the Australian Research Council.
RHM
thanks the Aspen Center for Physics for hospitality.

%\newpage
\vskip 3.0cm
\end{multicols}

{\onecolumn

\begin{table}
\caption{Organic superconductors described by our theory. 
All materials (except
the last one) have an average of half a hole per donor molecule. This
corresponds to a quarter-filled band. T$_c$ is the superconducting
transition temperature at the given pressure. Five of the materials 
have the unusual property that
the resistivity, $\rho$, decreases with increasing temperature 
above $T_c$
suggesting a direct transition from an insulating
phase into a superconducting phase ($d \rho/d T < 0$). 
There is evidence of charge ordering (CO) and/or an insulating phase
(with a metal-insulator transition temperature T$_{MI}$) in close 
proximity to the superconducting phase.
$\theta$-(BEDT-TTF)$_2$ I$_3$ is close to a charge ordered
insulator as when the 
anion, I$_3$, is replaced with RbZn(SCN)$_4$, CsZn(SCN)$_4$,
CsZn(SCN)$_4$ or CsCo(SCN)$_4$, the material becomes            
a charge-ordered insulator \protect\cite{MoriH}.
(Y=(C$_2$O$_4$)$_3$$ \cdot$PhCN).
}
%\vskip0.5cm
\label{table1}
\begin{tabular}{lllllllll}
Material & Pressure & T$_c$(K) & Ref. & ${d \rho \over d T} < 0$ & Pressure 
& CO & T$_{MI}(K)$ & Ref. \\
\hline
$\theta$-(BEDT-TTF)$_2$I$_3$ & 1 bar  & 3.6  &\protect{\onlinecite{Kobayashi}}& No &  &  &  \\
$\beta$''-(BEDT-TTF)$_2$ SF$_5$CH$_2$CF$_2$SO$_3$ &  1 bar  &  5.2 &\protect
{\onlinecite{Geiser}} & No & 1 bar & Yes  &  &\protect{\onlinecite{Schlueter}} \\
(BEDT-TTF)$_2$ReO$_4$ & 4 kbar & 2  &\protect\onlinecite{Parkin} 
& No& 1 bar &   Yes &   77  &\protect\onlinecite{Ravy}   \\    
$\beta$''-(BEDT-TTF)$_4$Pd(CN)$_4$H$_2$O & 7.0 kbar &1.2&\protect{\onlinecite{MoriT}}& Yes & 1 bar  &  & 70 &\protect{\onlinecite{MoriT}}\\
$\beta$''-(BEDT-TTF)$_4$Pt(CN)$_4$H$_2$O & 6.5 kbar & 2.0 &\protect{\onlinecite{MoriH}}& Yes & 
1 bar  &    & 120  &\protect{\onlinecite{MoriH}}\\
%$\beta$''-(BEDT-TTF)$_4$H$_3$OFe(C$_2$O$_4$)$_3$$ \cdot$PhCN
$\beta$''-(BEDT-TTF)$_4$H$_3$OFeY
& 1 bar &8.5&\protect{\onlinecite{Kurmoo}} & No & 1 bar &  Yes &  &\protect{\onlinecite{Martin}}  \\ 
$\beta$''-(BEDT-TTF)$_4$H$_3$OCrY& 1 bar & 
 5.5 &  \protect{\onlinecite{Kurmoo}} & No &  1 bar &  Yes &  &\protect{\onlinecite{Martin}} \\    
$\theta$-(BETS)$_2$(Cl$_2$ TCNQ) & 3.5 kbar & 1.3  & \protect{\onlinecite{Kondo}} & Yes & 8.5 kbar  &   & 22  & \protect{\onlinecite{Kondo}} \\
$\beta$''-(BEDO-TTF)$_2$ReO$_4$ $\cdot$ H$_2$O & 1 bar & 2-3  & \protect\onlinecite{Beckmann} & Yes &   &   &   &  \\                      
$\theta$-(DIETS)$_2$Au(CN)$_4$ & 10 kbar uniaxial & 8  & 
\protect{\onlinecite{imakubo}} & No  & 1 bar  &   & 220  &
 \protect{\onlinecite{imakubo2}} \\
$\beta$''-(BEDT-TTF)$_3$Cl$_2$ $\cdot$ 2H$_2$O&  16 kbar  & 2-3 &
\protect{\onlinecite{Lubczynski}}& Yes &1 bar  & Yes & 150  &
\protect{\onlinecite{Gaultier}} \\    
\end{tabular}
\end{table}
}

\begin{references}
\bibitem{note} This is in contrast to Ref. \onlinecite{scalapino}
where the charge-density wave instability that produces
the charge fluctuations responsible for $d_{xy}$ pairing
is due to Fermi surface nesting.
\bibitem{theta}
%Quantum chemistry calculations using the
Extended H\"uckel calculations
suggest that this is the appropriate
tight-binding model for 
$\theta$- type crystal structures.
 See Reference \protect\cite{mckenzie} for details.
\end{references}
\end{document}